\documentclass[11pt,a4paper]{article}
\usepackage{cite}
\usepackage{multicol,bbm}
\usepackage{amsmath, amssymb, amsfonts}
\usepackage{graphicx}
\usepackage{caption}
\usepackage{subcaption}
\usepackage{color}
\usepackage[utf8]{inputenc}
\usepackage{float}
\makeatletter \@addtoreset{equation}{section} \makeatother
\begin{document}
\title{Dirac field in the background of a planar defect}
\author{D. Bazeia$^1$\thanks{bazeia@fisica.ufpb.br}\ \ and A. Mohammadi$^2$\thanks{azadeh.mohammadi@df.ufpe.br}\\
\textit{$^1$Departamento de F\'{\i}sica, Universidade Federal da Para\'{\i}ba}\\\textit{58051-900 Jo\~ao Pessoa, Para\'{\i}ba, Brazil}\vspace{0.3cm}\\\textit{$^2$Departamento de F\'isica, Universidade Federal de Pernambuco}\\\textit{52171-900 Recife, Pernambuco, Brazil}}

\maketitle

\begin{abstract}
We study massless Dirac fermions in the background of a specific planar topologically nontrivial configuration in the
three-dimensional spacetime. The results show the presence of massive bound states, phase shifts and the consequent differential cross section for the scattering of fermions in the weak coupling regime. Despite the nontrivial topology of the background field, no fermionic zero mode is found.
\end{abstract}

\maketitle

\section{Introduction}

In general, the interaction of fermion fields with solitonic backgrounds, topological or nontopological, may create or affect various interesting physical phenomena like charge and fermion number fractionalization, vacuum polarization and Casimir effect, superconductivity, Bose-Einstein condensation, conducting polymers and localization of fermions in the braneworld scenarios (see for example \cite{fractionalization,condpolymer1,condpolymer2,casimir1,casimir2,supercond,hadron1,hadron2,2Dmat1,BEcond1,BEcond2,branelocalization1,branelocalization2}).  Besides that, there are interesting works related to the investigation of fermions in soliton backgrounds in the context of supersymmetry (see for example \cite{Witten,Dionisio,Shifman,Brihaye}).
The massless Dirac fermions emerge as the quasiparticles in various novel materials such as graphene and topological insulators exhibiting intriguing behaviors \cite{Neto,Qi}. In the context of 2D materials like graphene, it is important to study the band structure and properties of the trapped Dirac electron states and the consequent electronic properties of the material in the presence of a defect (see for instance \cite{2Dmat1,2Dmat2,2Dmat3}).

In $2+1$ dimensions there are two particularly intresting types of solitons; vortices appearing in Maxwell-Higgs and Chern-Simons theories where one can attribute electric charge to the vortex in the latter case. 
For specific choices of the Higgs or scalar potential, the minimum energy static vortex solutions satisfy a set of first-order differential self-duality equations, 
or known as Bogomol'nyi equations. In \cite{main1} the authors introduced a new set of topological defects respecting self-duality condition in $2+1$ dimensions where the translational symmetry of the system is broken. In the models they considered, there is only one real scalar field which in general makes it impossible to have topological defects thanks to Derrick-Hobart no go theorem. The key point in their work to circumvent the obstruction was to introduce explicit space dependence in the potential term of the boson field. We think it would be intresting to study fermions in detail, considering the effect of symmetries/symmetry breakings, in a specific 2D nontrivial configuration with this characteristic.

The fermionic zero modes are relevant to the quantum theory of the models, while the zero modes of the bosonic fluctuations determine the collective coordinates that describe the solitons. However, in supersymmetric models, the fermionic zero modes can be directly related to the zero modes of bosonic fluctuations describing massless modes around the vortices. 
Fermionic zero modes in the Dirac equation for fermions coupled to a topologically nontrivial defect background are important in systems belonging to a large domain in physics, going from high energy to condensed matter physics. Specially their relation with the topology of the background defect is of considerable interest. In \cite{JRzeromode}, Jackiw and Rossi showed that the Dirac field has $|n|$ zero modes in the $n$-vortex background field. This means that the fermionic zero modes are protected due to the nontrivial topology of the background soliton. In this line of work one can find large number of papers in the literature (see for example \cite{zeromode} and references therein). However, we discuss here a counter example when the system does not respect translational symmetry and the Lagrangian has explicit space dependence. Besides that the model does not contain a gauge field and the corresponding topological flux associated to the vorticity of the system. We show that in the model considered in this paper, there is no fermionic zero mode, although the background configuration produces a planar topological structure that can be used to simulate a skyrmion-like structure with unity skyrmion number, a subject explored before in Refs.~\cite{main2,main3}.

We studied fermions in the background of several 1D kinklike configurations in \cite{A,B} and found the fermionic zero mode as well as all other massive bound spectrum. In the current paper, we consider a massless Dirac field in a specific rotationally symmetric and localized topological structure in 2+1 dimensions where the system does not respect translational symmetry. We are interested in the fermionic bound energy spectrum as well as the scattering phase shift due to the interaction with the defect. When the coupling constant of the fermion-soliton interaction is small compared to the self-interaction of the boson field, resulting in a heavy soliton compared to the scales appearing in the system, it is possible to ignore the effect of the fermion on the soliton which is the case in this paper. In this sense we say the defect is the background perturbation for the Dirac field. Due to the presence of a solitonic background as trapping potential, the fermion field spectrum can be distorted, i.e., bound states can appear and continuum states can change as compared with the free fermion.

The work deals mainly with the fermionic bound energy spectrum besides the scattering phase shift in the presence of the background defect studied before in \cite{main1,main2,main3}. In this model we can see that there is no fermionic zero mode, although the background configuration is topologically nontrivial. In Sec.~\ref{sec:model} we introduce the theory and discuss about the symmetries of the system, which we use to drive the simplified versions of the equations of motion. In Sec.~\ref{first: model} we briefly review a model that can be used to describe a skyrmion-like structure with unity skyrmion number. In the model to be considered here it is easy to see that there is no need to add a gauge field to have a well-defined theory, in contrast with the Maxwell-Higgs vortex model. Finally, in Sec.~\ref{sec:end} we summarize and discuss the main results of the current work.

\section{Yukawa coupling}
\label{sec:model}
The Lagrangian density adopted in the present work has the following form
\begin{align}\label{a6}
	\mathcal{L} = \bar{\psi}\ i\gamma^\mu\partial_\mu \psi  - g \phi \, \bar{\psi}\psi+\frac{1}{2} \partial_\mu \phi\partial^\mu \phi -U(r;\phi),
\end{align}  
where $\psi$ and $\phi$ are fermion and boson fields, respectively. We work in $2+1$ dimensions and write the Lagrangian density in the form
\begin{align}\label{model}
	\mathcal{L} = \mathcal{L}_b+\mathcal{L}_f,
\end{align}
where the bosonic contribution is given by \cite{main1}
\begin{align}\label{modelb}
	\mathcal{L}_b = \frac{1}{2} \partial_\mu\phi\partial^\mu \phi
	- \frac{1}{r^2}V(\phi),
\end{align}
and
\begin{align}\label{modelf}
	\mathcal{L}_f = \bar{\psi}\,i\gamma^\mu\partial_\mu\psi
	- g \phi\,\bar{\psi}\psi.
\end{align}
In this paper, we are interested in studying the fermion system given by the Lagrangian density $\mathcal{L}_f$ interacting with the background planar defect configuration, the solution of the equation of motion considering $\mathcal{L}_b$ with the scalar potential
\begin{align}\label{scalar-pot-V}
V(\phi)=\frac{a}{2} \, (v^2-\phi^2)^2.
\end{align}
The fermion field couples to the bosonic structure via the Yukawa coupling parameter $g$, and the equation of motion, considering the Lagrangian
$\mathcal{L}_f$, has the form
\begin{align}\label{e-o-m}
\left(i\gamma^\mu\partial_\mu - g \phi \right)\psi=0 \ .
\end{align}
As it is clear from the equation of motion (\ref{e-o-m}) the system breaks parity symmetry which is not unusual in $2+1$ dimensions, due to the fact that the  parity symmetry acts differently and parity transformation should be taken as reflection in just one of the spatial axes. Besides that, the system does not have energy-reflection symmetry. Therefore, we do not expect symmetric energy spectrum around the line $E=0$. The representation we choose for the Dirac matrices is $\gamma^0 = \sigma_3$, $\gamma^1 = i \sigma_2 $ and $\gamma^2 = -i \sigma_1$. The charge-conjugation transformation is representation dependent. The system is symmetric under this transformation and in this specific representation the charge-conjugation operator is $\gamma^2$. 

In $2+1$ dimensions, the mass dimension of the bosonic field $\phi$, the spinor field $\psi$ and the coupling constant $g$ are $1/2$, $1$ and $1/2$, respectively. We rescale all mass scales by the value of the field $\phi$ at infinity as $\phi\rightarrow \phi/ v$, $\psi\rightarrow \psi/ v^2$, $r\rightarrow r v^2$, $g\rightarrow g/ v$ and $a\rightarrow a v^2$. Therefore, from now on all parameters of the system are dimensionless. 

We then define  
\begin{equation}
\psi \equiv e^{-i E t} \begin{pmatrix}
		\psi_1(r,\theta)  \\
		\psi_2 (r,\theta)
	   \end{pmatrix}
\end{equation}	
in order to get the Dirac equation in the explicit form
\begin{align}\label{e_o_m_comp}
     \left(i e^{-i \theta}\partial_r+\frac{e^{-i \theta}}{r}\partial_\theta\right)\psi_2(r,\theta) &= -\left[E-g \phi(r)\right]\psi_1(r,\theta),\nonumber\\
\left(i e^{i \theta}\partial_r-\frac{e^{i \theta}}{r}\partial_\theta\right)\psi_1(r,\theta) &= -\left[E+g \phi(r)\right]\psi_2(r,\theta).
\end{align}
The rotation symmetry allows us to write down
an ansatz for the solution to the Dirac equation using separation of variables
\begin{align}\label{spinor}
\psi_1(r,\theta)=\psi_1(r)e^{i(j-1/2)\theta},\nonumber\\
\psi_2(r,\theta)=\psi_2(r)e^{i(j+1/2)\theta},
\end{align}
where $\psi_1(r)$ and $\psi_2(r)$ are complex in general. Substituting the above relations in the equations of motion leads to
\begin{align}\label{e_o_m_radial}
     i\left(\partial_r+\frac{(j+1/2)}{r}\right)\psi_2(r) &= -\left[E-g \phi(r)\right]\psi_1(r),\nonumber\\
i\left(\partial_r-\frac{(j-1/2)}{r}\right)\psi_1(r) &= -\left[E+g \phi(r)\right[\psi_2(r).
\end{align}
This set of equations are not symmetric under $j\rightarrow -j$ which is reflecting the fact that the system does not respect parity. Separating imaginary and real parts of the components of the spinor field as 
\begin{align}\label{spinor-re-im}
\psi_1(r)=\psi_1^R(r)+i\psi_1^I(r),\nonumber\\
\psi_2(r)=\psi_2^R(r)+i\psi_2^I(r),
\end{align}
results in
\begin{align}\label{e_o_m_radial_re_im1}
     \left(\partial_r+\frac{(j+1/2)}{r}\right)\psi_2^I(r) &= \left(E-g \phi(r)\right)\psi_1^R(r),\nonumber\\
\left(\partial_r-\frac{(j-1/2)}{r}\right)\psi_1^R(r) &= -\left(E+g \phi(r)\right)\psi_2^I(r),
\end{align}
and
\begin{align}\label{e_o_m_radial_re_im2}
     \left(\partial_r+\frac{(j+1/2)}{r}\right)\psi_2^R(r) &= -\left(E-g \phi(r)\right)\psi_1^I(r),\nonumber\\
\left(\partial_r-\frac{(j-1/2)}{r}\right)\psi_1^I(r) &= \left(E+g \phi(r)\right)\psi_2^R(r).
\end{align}
It is enough to solve the first  set of equations (\ref{e_o_m_radial_re_im1}), because of the symmetry under $\psi_1^I(r) \to \psi_1^R(r)$ and $\psi_2^R(r) \to -\psi_2^I(r)$. The decoupled equations are
\begin{align}\label{sec-order-1}
\psi_1''+\left[\frac{1}{r}-\frac{g \phi'}{(E+g\phi)}\right]\psi_1'+\left[-\frac{(j-1/2)^2}{r^2}+\frac{(j-1/2)}{r} \frac{g \phi'}{(E+g\phi)}+(E^2-g^2\phi^2)\right]\psi_1=0
\end{align}
and
\begin{align}\label{sec-order-2}
\psi_2''+\left[\frac{1}{r}+\frac{g \phi'}{(E-g\phi)}\right]\psi_2'+\left[-\frac{(j+1/2)^2}{r^2}+\frac{(j+1/2)}{r} \frac{g \phi'}{(E-g\phi)}+(E^2-g^2\phi^2)\right]\psi_2=0
\end{align}
where we have dropped the indices $R$ and $I$. Making the change of variables $\chi_1\equiv\sqrt{r}\ \psi_1$ and $\chi_2\equiv\sqrt{r}\ \psi_2$, we have
\begin{align}\label{sec-order-3}
&\chi_1''-\frac{g \phi'}{(E+g\phi)}\chi_1'+\left[-\frac{(j-1/2)^2}{r^2}+\frac{(j-1/2)}{r} \frac{g \phi'}{(E+g\phi)}+(E^2-g^2\phi^2)+\frac{1}{4r^2}\right.\nonumber\\
&+\left. \frac{g \phi'}{2r(E+g\phi)}\right]\chi_1=0
\end{align}
and
\begin{align}\label{sec-order-4}
&\chi_2''+\frac{g \phi'}{(E-g\phi)}\chi_2'+\left[-\frac{(j+1/2)^2}{r^2}+\frac{(j+1/2)}{r} \frac{g \phi'}{(E-g\phi)}+(E^2-g^2\phi^2)+\frac{1}{4r^2}\right.\nonumber\\
&-\left.\frac{g \phi'}{2r(E-g\phi)}\right]\chi_2=0
\end{align}
After some manipulations, we can get rid of the first derivative terms in the above equations for $\chi_1$ and $\chi_2$ which results in 
\begin{align}\label{sec-order-5}
&\frac{d^2\chi_1}{d\rho_1^2}+\left[-\frac{(j-1/2)^2}{r^2(\rho_1)}+\frac{(j-1/2)}{r(\rho_1)} \frac{g \phi'}{(E+g\phi)}+(E^2-g^2\phi^2)+\frac{1}{4r^2(\rho_1)}\right.\nonumber\\
&\left.+\frac{g \phi'}{2r(\rho_1)(E+g\phi)}+\frac{g \phi''}{2(E+g\phi)}-\frac{3g^2 \phi'^2}{4(E+g\phi)^2}\right]\chi_1=0
\end{align}
and
\begin{align}\label{sec-order-6}
&\frac{d^2\chi_2}{d\rho_2^2}+\left[-\frac{(j+1/2)^2}{r^2(\rho_2)}+\frac{(j+1/2)}{r(\rho_2)} \frac{g \phi'}{(E-g\phi)}+(E^2-g^2\phi^2)+\frac{1}{4r^2(\rho_2)}\right.\nonumber\\
&\left.-\frac{g \phi'}{2r(\rho_2)(E-g\phi)}-\frac{g \phi''}{2(E-g\phi)}-\frac{3g^2 \phi'^2}{4(E-g\phi)^2}\right]\chi_2=0
\end{align}
where the prime shows the derivative with respect to $r$ and besides that we have used the relation $(\partial_r+f/2)(\partial_r+f/2)=\partial_r^2+f\partial_r+\partial_rf/2+f^2/4\equiv \partial_\rho^2$. To find the phase shift we need only to study the behavior of the solutions for  large distances. Therefore, we can expand the functions for large $r$ \cite{planarBH}. At this point we need to know the behavior of the defect solution which is driven by the field $\phi$, so we turn attention to the bosonic sector of the model to proceed.
\\

\section{Planar defect configuration}
\label{first: model}

We review here briefly the bosonic model which appears in Eq.~\eqref{a6} with the following dimensionless scalar potential \cite{main1,main2}
\begin{align}\label{scalar-pot-1}
U(r;\phi)=\frac{1}{2 r^2} \, (1-\phi^2)^2 ,
\end{align}
where we have chosen the rescaled parameter $a=1$. Considering the explicit spatial dependence in the potential term which is the case here one can avoid \cite{main1} the Derrick no go theorem which states that models described by a single real scalar field cannot support topological defects in more than one spatial dimensions. 

\begin{figure}[h]
\begin{center}
\boxed{	{\includegraphics[width=0.5\textwidth]{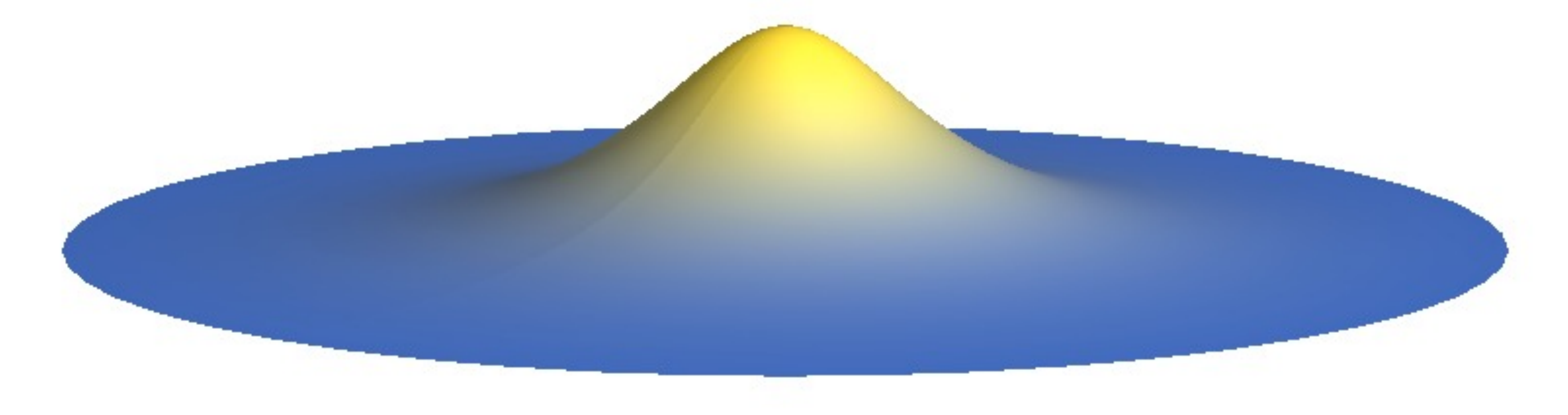}}}
\caption{An illustration of the defect configuration that appear from Eq.~\eqref{sol-eomscalar1} in Cartesian coordinate, considering the positive sign solution.}
    \label{fig0}
\end{center}
\end{figure}

The equation of motion related to the Lagrangian $\mathcal{L}_b$ reads
\begin{align}\label{eomscalar1}
r^2 \frac{d^2 \phi}{dr^2}+r\frac{d \phi}{dr}+2 \phi (1-\phi^2)=0.
\end{align}
Analytical solutions of the above equation of motion are
\begin{align}\label{sol-eomscalar1}
\phi(r)=\pm \frac{1-r^{2}}{1+r^{2}},
\end{align}
which is illustrated in Fig.~\ref{fig0} choosing the positive sign. The energy of the static solutions identifies the mass associated to them, which is given by $M= {8\pi}/{3}$.
The solutions are linearly stable, and the linear stability analysis of the model results in the zero mode $\eta(r)=N {r^{2}}/{(1+r^{2})^2}$,
where $N$ is the normalization factor \cite{main2}. In the current work we consider the potential \eqref{scalar-pot-1} and the solution \eqref{sol-eomscalar1} with the positive sign.

We are interested in studying the fermion field in the background of the above topological defect. We are working with dimensionless parameters, and it is easy to see that the fermionic threshold energies are $E=\pm g$. Therefore, all bound states should appear between these two limiting values.

\begin{figure}[h]
\begin{center}
	{\includegraphics[width=0.45\textwidth]{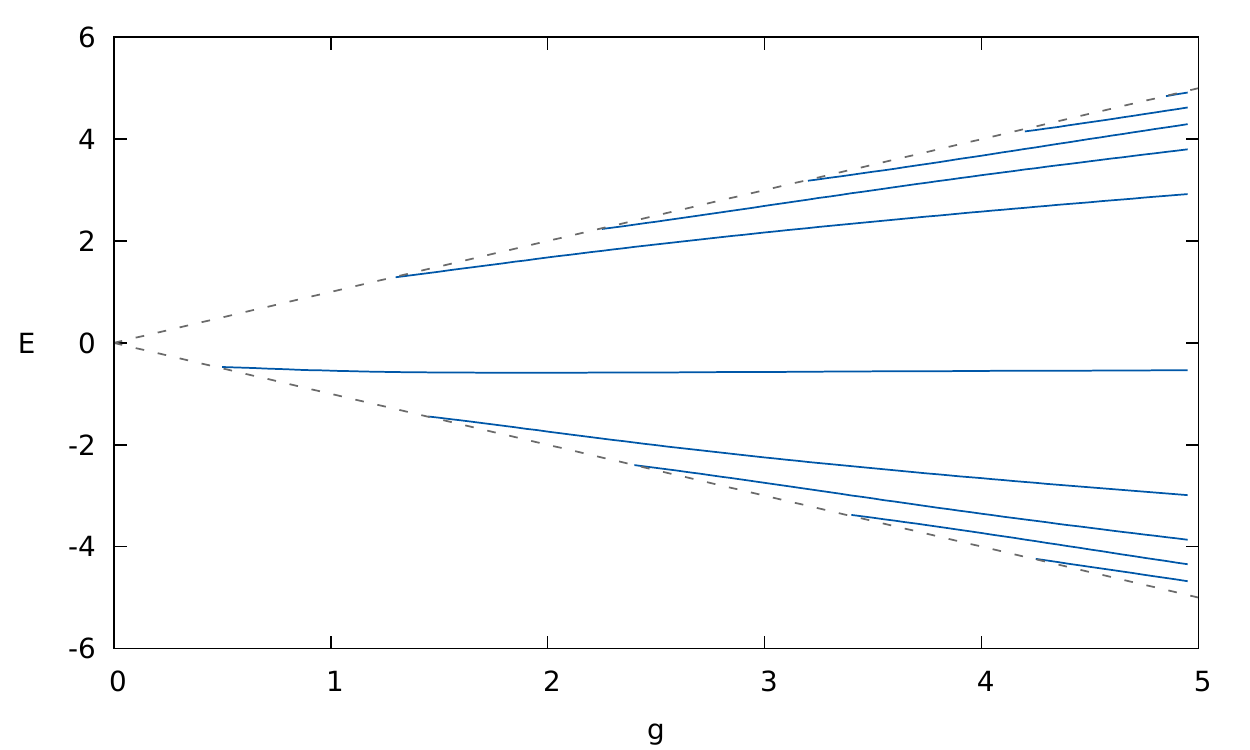}}
	{\includegraphics[width=0.45\textwidth]{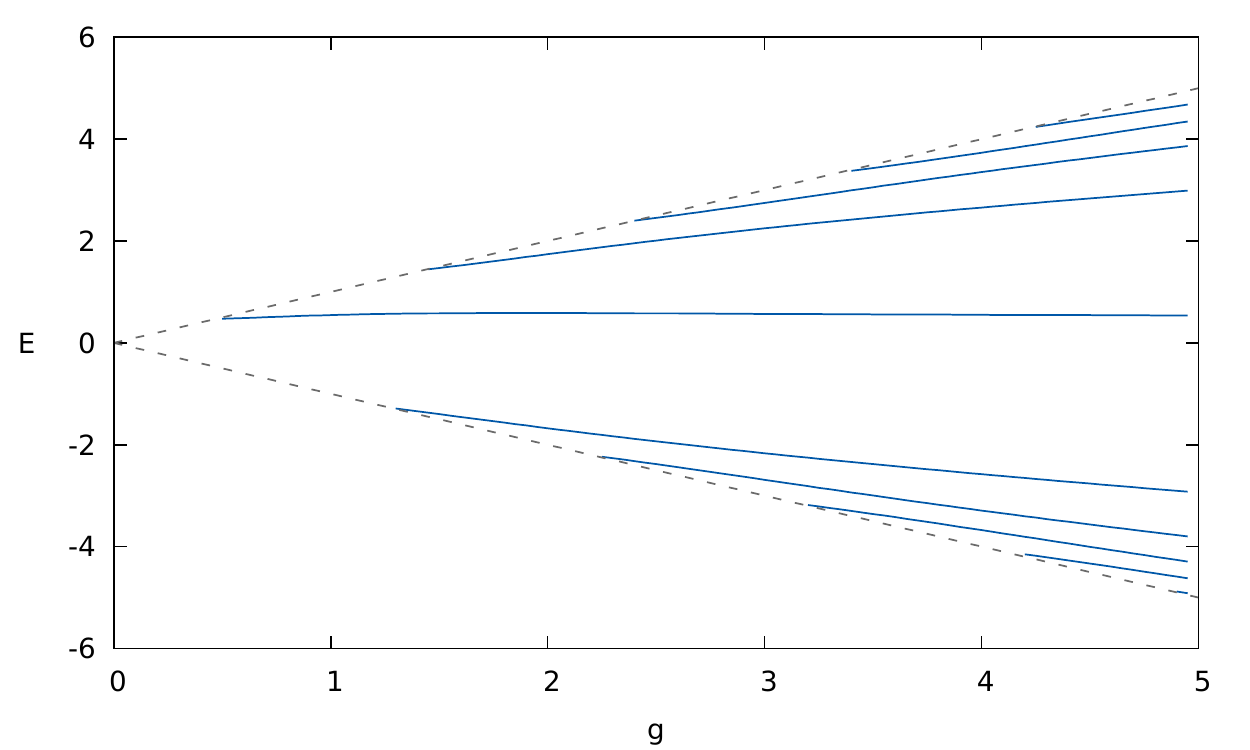}}
	\caption{The fermion bound energy spectrum as a function of the coupling $g$ for $j=0.5$ (left) and $j=-0.5$ (right). The dashed lines identify the threshold energies.}
    \label{fig1}
\end{center}
\end{figure}

\begin{figure}[h]
\begin{center}
	{\includegraphics[width=0.5\textwidth]{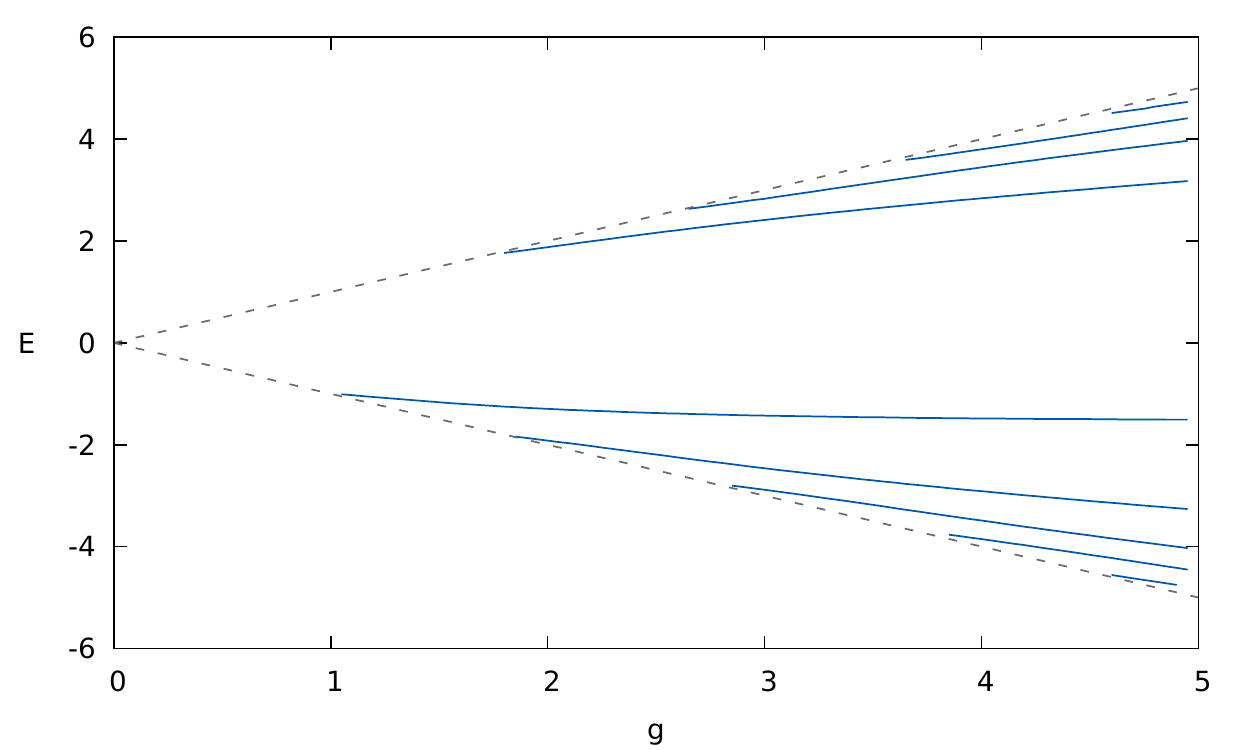}}
	\caption{The fermion bound energy spectrum as a function of the coupling $g$ for $j=1.5$. The dashed lines identify the threshold energies.}
    \label{fig2}
\end{center}
\end{figure}

As one can see in Fig.~\ref{fig1} and Fig.~\ref{fig2}, which show the fermion bound state spectrum as a function of the coupling $g$ for three values of $j=\pm0.5, 1.5$, there is no fermionic zero mode which means the topology of the system does not protect these modes. Besides that, one can see the energy spectrum is not symmetric for positive and negative bound energies thanks to the fact that the system does not respect energy-reflection symmetry. Also, the bound spectrum as a function of $g$ for $j=\pm0.5$ shows that it is not symmetric under $j\rightarrow-j$, thanks to the parity symmetry breaking.

Another information that appears in Figs.~\ref{fig1} and \ref{fig2} is that there is no massive bound state for small values of $g$. We may then say that in the weak coupling regime, no fermion bound state can be attached to such a planar topological structure, in contrast with the case of vortices in the Maxwell-Higgs model investigated in Ref.~\cite{JRzeromode}.  

Let us now look at the scattering states and the phase shift appearing due to the presence of the topological structure. Substituting the solution (\ref {sol-eomscalar1}) in the second order equations (\ref {sec-order-5}) and (\ref {sec-order-6}) we obtain
\begin{align}\label{sec-order-7}
&\frac{d^2\chi_1}{d\rho_1^2}+\left[(E^2-g^2)+\frac{1/4-\bar m^2}{\rho_1^2}+\mathcal{O}[1/\rho_1^4]\right]\chi_1=0,
\end{align}
and
\begin{align}\label{sec-order-8}
&\frac{d^2\chi_2}{d\rho_2^2}+\left[(E^2-g^2)+\frac{1/4-\bar n^2}{\rho_2^2}+\mathcal{O}[1/\rho_2^4]\right]\chi_2=0,
\end{align}
where $\rho_1=r+\frac{2g}{(E-g)r}+\mathcal{O}[1/r^3]$, $\rho_2=r-\frac{2g}{(E+g)r}+\mathcal{O}[1/r^3]$, $n\equiv j+1/2$, $m\equiv n-1$, $\bar n^2\equiv n^2-4g^2$ and $\bar m^2\equiv m^2-4g^2$. As a result, for large r, ignoring the terms $\mathcal{O}[1/\rho_1^4]$ and $\mathcal{O}[1/\rho_2^4]$, we can obtain analytic solutions
\begin{align}\label{sol-large r}
\chi_1=J_{\bar m}(\lambda\rho_1),\nonumber\\
\chi_2=J_{\bar n}(\lambda\rho_2),
\end{align}
where $\lambda\equiv (E^2-g^2)$. The above solutions converge to
\begin{align}\label{sol-inf r}
\chi_1=J_{\bar m}(\lambda r),\nonumber\\
\chi_2=J_{\bar n}(\lambda r),
\end{align}
in the limit $r\rightarrow \infty$.
Now, we can compare this result with the one in the absence of the soliton where we obtain the following approximate expressions for the phase shift for the upper and lower components of the spinor field
\begin{align}\label{phase-shift}
\delta_u&=\frac{\pi}{2}(m-\bar m)\nonumber\\
\delta_d&=\frac{\pi}{2}(n-\bar n)
\end{align}
For small values of $g$, one can expand the above expressions and write 
\begin{align} \label{phase-shift-expansion}
\delta_u &=\textrm sgn(m)\left[\frac{\pi g^2}{m}+\mathcal{O}[{1}/{m^3}]\right]\\
\delta_d &=\textrm sgn(n)\left[\frac{\pi g^2}{n}+\mathcal{O}[{1}/{n^3}]\right] 
\end{align}
We can use this to calculate the differential scattering cross section which is given by
\begin{align}
\frac{d\sigma}{d\theta}=|f_{E}(\theta)|^2 \ ,
\end{align}
where
\begin{align}
f_{E}(\theta)=\sqrt{\frac{1}{2i\pi \lambda}}\sum_{m=-\infty}^{+\infty}(e^{2i\delta^{(m)}}-1)e^{i m \theta},
\end{align}
for each component. Considering small values of $g$, in the weak coupling regime the main contribution comes from the term with $m=0$, and for
$\delta^{(0)}$ one needs to use the results of eq.~(\ref{phase-shift}). Therefore, the total differential scattering cross section for a massless fermion in this system can be approximated as
\begin{align}
\frac{d\sigma}{d\theta}=&\frac{1}{2}\left(\frac{d\sigma_u}{d\theta}+\frac{d\sigma_d}{d\theta}\right)\approx\frac{(e^{2\pi g}-1)^2}{2\pi\lambda} \approx \frac{2 \pi g^2}{E^2} .
\label{tot-cross-section}
\end{align} 
It is easy to see that in the limit $g\rightarrow 0$, where the interaction with the topological structure disappears, the scattering differential cross section goes to zero, as expected.

\section{Comments and Conclusions}
\label{sec:end}

In this work, we have investigated the Yukawa interaction between a massless Dirac fermion and a boson field capable of generating a planar topological structure as the one constructed in Refs.~\cite{main1,main2,main3}.  We have presented all its bound state solutions energies numerically and studied analytically the scattering phase shift and the resulting differential cross section due to a nontrivial defect configuration in the weak coupling regime. 

The investigation had started with the fermionic model interacting with the aforementioned topological defect configuration considering the defect as background. Then, we have briefly reviewed the bosonic sector of the model describing the background defect. We have used the model to calculate the fermion bound states as a function of the Yukawa coupling parameter, and we noticed the absence of fermionic zero modes. We have shown that the fermionic energy spectrum is not symmetric around the line $E=0$ thanks to the fact that the system breaks the energy-reflection symmetry. We also calculated the phase shifts, and for small values of the Yukawa parameter $g$ we could find a closed expression for the differential cross section that describes the scattering of fermions from the localized topological structure. We have confirmed that in the absence of the interaction term between the fermion and the topological structure (the limit $g\rightarrow 0$) the scattering differential cross section goes to zero, which is the expected result.
In the case of small $g$, we see from the results displayed in Figs.~\ref{fig1} and \ref{fig2}, that no fermion bound state is present, so we are only left with scattering states. One can conclude that at least in the weak coupling regime, the planar structure does not acquire fermionic bound states, in contrast with the behavior of vortices in the Maxwell-Higgs model reported before in Ref.~\cite{JRzeromode}. 
 
\section*{Acknowledgments}
DB would like to thank the Brazilian agency CNPq for partial financial support, under the contract 306614/2014-6. AM acknowledges Universidade Federal de Pernambuco for the partial financial support, under grant {\it Qualis A}. 


\end{document}